\documentclass[aps,prl,twocolumn,showpacs]{revtex4}
\usepackage{amssymb}
\usepackage{amsmath, amsthm}

\newcommand{\be}{\begin{equation}}
\newcommand{\ee}{\end{equation}}

\begin{document}

% Use the \preprint command to place your local institutional report
% number in the upper righthand corner of the title page in preprint mode.
% Multiple \preprint commands are allowed.
% Use the 'preprintnumbers' class option to override journal defaults
% to display numbers if necessary
%\preprint{}

\title{Breakdown of the initial value formulation of 
scalar-tensor gravity and its physical meaning}

% repeat the \author .. \affiliation  etc. as needed
% \email, \thanks, \homepage, \altaffiliation all apply to the current
% author. Explanatory text should go in the []'s, actual e-mail
% address or url should go in the {}'s for \email and \homepage.
% Please use the appropriate macro for each each type of information

% \affiliation command applies to all authors since the last
% \affiliation command. The \affiliation command should follow the
% other information
% \affiliation can be followed by \email, \homepage, \thanks as well.
\author{Valerio Faraoni}
\email[]{vfaraoni@ubishops.ca}
\author{Nicolas Lanahan-Tremblay}
\email[]{ntremblay@ubishops.ca}
%\homepage[]{Your web page}
%\thanks{}
%\altaffiliation{}
\affiliation{Physics Department, Bishop's University\\
Sherbrooke, Qu\'{e}bec,  Canada J1M~1Z7}

%Collaboration name if desired (requires use of superscriptaddress
%option in \documentclass). \noaffiliation is required (may also be
%used with the \author command).
%\collaboration can be followed by \email, \homepage, \thanks as well.
%\collaboration{}
%\noaffiliation

\date{\today}

\begin{abstract}
We revisit singularities of two distinct kinds in the Cauchy problem of 
general scalar-tensor theories of gravity (previously discussed in the 
literature), and of metric and Palatini $f(R)$ 
gravity, in both their Jordan and 
Einstein frame representations. Examples and toy models are used to shed 
light onto the problem and it is shown that, contrary to common lore, the 
two conformal frames are equivalent with respect to the initial value 
problem.
\end{abstract}

% insert suggested PACS numbers in braces on next line
% insert suggested keywords - APS authors don't need to do this

%\maketitle must follow title, authors, abstract, \pacs, and \keywords
\maketitle

%\section{}
% Put \label in argument of \section for cross-referencing
%\section{\label{}}
%\subsection{}
%\subsubsection{}
% If in two-column mode, this environment will change to single-column
% format so that long equations can be displayed. Use
% sparingly.
%\begin{widetext}
% put long equation here
%\end{widetext}

%\def\theequation{\arabic{section}.\arabic{equation}}

\section{\label{sec:intro}I.~Introduction}

The 1998 discovery  of the acceleration of the cosmic 
expansion, obtained by studying type Ia supernovae \cite{SN}, 
spurred an enormous amount of activity on dark energy models, 
mostly based on cosmological scalar fields. Certain models 
are set in the context of scalar-tensor gravity instead of Einstein's 
theory, and are dubbed ``extended quintessence'' 
\cite{extendedquintessence}. Moreover, as an alternative to 
postulating a 
mysterious form of dark energy, various authors (\cite{CCT, 
CDDT}, see \cite{review} for a review) have considered the 
possibility that the acceleration of 
the universe is caused instead by a modification of gravity at 
large scales: the Einstein-Hilbert action 
\be\label{1}
S_{EH}=\frac{1}{2\kappa}\int d^4x \, \sqrt{-g} \, R 
+S^{(m)}\left[ g_{ab}, \psi \right] 
\ee
is generalized to
\be \label{2}
S=\frac{1}{2\kappa}\int d^4x \, \sqrt{-g} \, f(R) 
+ S^{(m)}\left[ g_{ab}, \psi \right]  \;,
\ee
where $f(R)$ is an arbitrary, twice differentiable, 
function of $R$. 
Here $\kappa \equiv 8\pi G$, $G$ is Newton's constant (that 
will be unity, together with the speed of light, in the 
geometrized units employed), $R$ 
is the Ricci curvature, $S^{(m)}=\int d^4x \, \sqrt{-g} \, {\cal 
L}^{(m)} \left[ g_{ab}, \psi \right] $ is the matter part of 
the  action, $\psi$ collectively denotes the matter fields, and 
we follow the notations of \cite{Wald}.

If the 
action~(\ref{2}) is varied with respect to the  metric 
$g_{ab}$, one obtains the {\em metric formalism} with fourth 
order field equations 
\cite{CCT,CDDT}; if the metric 
and the  connection $\Gamma^a_{bc}$ are considered as  
independent variables ({\em i.e.}, the connection is not the 
metric connection of $g_{ab}$), but the matter part of the 
action $S^{(m)}$ does not depend explicitly on $\Gamma$, one 
obtains the {\em Palatini formalism} with second 
order field equations \cite{Vollick}. 
If, instead, $S^{(m)}$  depends on $\Gamma$, one obtains {\em 
metric-affine} gravity \cite{SotiriouLiberati}.

It has been shown \cite{STequivalence} that metric $f(R)$ 
gravity is dynamically equivalent to a Brans-Dicke (BD) theory 
with BD parameter $\omega_0=0$, while Palatini $f(R)$ gravity is 
equivalent to an $\omega_0=-3/2$ BD theory. The general form of 
the scalar-tensor action, of which BD theory \cite{BD, Dicke} 
is the prototype, is \cite{ST} 
\begin{eqnarray} 
S_{ST} & = &\int d^4x \, \sqrt{-g} \, \left[ \frac{f(\phi)R}{2} 
-\frac{\omega(\phi) }{2}\, \nabla^c\phi \nabla_c\phi -V(\phi) 
\right] \nonumber \\
& +& S^{(m)} \left[ g_{ab}, \psi \right] \;,\label{3}
\end{eqnarray}
where $\phi$ is the BD-like scalar field and 
$f(\phi)>0$ is required in order for the effective 
gravitational coupling to be positive and the  graviton to 
carry positive kinetic energy and not being a ghost.  $V(\phi)$ 
is the scalar field potential, while $f(\phi)$ and 
$\omega(\phi)$ are two ({\em a priori} arbitrary) coupling 
functions. BD theory is recovered as the special case 
$f(\phi)=\phi$ and $\omega (\phi)=\omega_0/\phi$, with $\omega_0=$const. 
The  field 
equations derived from the action~(\ref{3}) are
\begin{widetext}
\begin{eqnarray}
&& f(\phi) \left( R_{ab}-\frac{1}{2} g_{ab} R \right)= 
\omega (\phi) \left( \nabla_a\phi\nabla_b \phi -\frac{1}{2} \, 
g_{ab} 
\nabla^c\phi\nabla_c \phi \right) -V g_{ab}+\nabla_a \nabla_b 
f -g_{ab}\Box f +T_{ab}^{(m)} \;,  \label{4} \\
&& \nonumber \\
&& \left[ \omega+\frac{3(f')^2}{2f} \right] \Box \phi + \left( 
\frac{\omega '}{2} +\frac{3f'f''}{2f}+\frac{ \omega f'}{2f} 
\right) \nabla_c\phi \nabla^c \phi   =\frac{f'}{2f}\, 
T  +2V'-\frac{2Vf'}{f} \;, \label{5}
\end{eqnarray}
\end{widetext}
where a prime denotes differentiation with respect to 
$\phi$, $\Box \equiv g^{ab}\nabla_a\nabla_b$, and 
$T_{ab}=\frac{-2}{ \sqrt{-g} }\, \frac{\delta S^{(m)}}{\delta 
g^{ab} }$.

The original motivation for BD theory was the 
implementation in relativistic gravity of 
the Mach principle, which is not fully embodied in general 
relativity, by promoting Newton's constant to 
the role of  a dynamical field determined by the environment   
\cite{BD,  Dicke}. Later on, it was discovered that string 
theories and  supergravity contain BD-like scalars: in fact, 
the low energy limit of  the bosonic string theory (which, 
although unphysical because it does not contain fermions and 
is not supersymmetric, was one of the early string theories) 
is indeed an $\omega_0=-1$ BD theory 
\cite{bosonicstring}.  Moreover, BD theory can be derived 
from  higher-dimensional Kaluza-Klein theory, higher 
dimensionality 
being an essential feature of all modern high energy theories. 
A $p$-brane model in $D$ dimensions leads, after 
compactification, to a BD theory with parameter \cite{Duffetal}
\be
\omega_0=-\, \frac{  (D-1)(p-1) -(p+1)^2}{(D-2)(p-1) -(p+1)^2} 
\;.
\ee
These properties have renewed the interest in BD and 
scalar-tensor gravity since the 1980s, following the rise of 
string theory.  However, the more recent surge of interest in 
scalar-tensor gravity that we are witnessing is motivated by 
cosmology and is linked to attempts to explain the present 
cosmic acceleration (see \cite{mybook, FujiiMaeda} for reviews 
of scalar-tensor gravity  in the cosmology of the early and 
present universe).

Motivated by the past and recent interest and also by 
developments in numerical relativity, the initial value 
problem of scalar-tensor gravity was studied   
by Salgado \cite{Salgado} who, using a first order 
hyperbolicity analysis, 
showed that the Cauchy problem is well-formulated for  
theories of the form
\begin{eqnarray}
S & = & \int d^4x \, \sqrt{-g}\, \left[ \frac{f(\phi) R}{2} 
-\frac{1}{2} 
\nabla^c\phi \nabla_c\phi -V(\phi)\right] \nonumber \\
&&\nonumber \\
&+& S^{(m)} 
\left[ g_{ab}, \psi \right] \label{6}
\end{eqnarray}
when $S^{(m)}$ is ``reasonable'' \footnote{Strictly speaking, a 
separate analysis is needed for each different form of matter, 
but it is expected that ``reasonable'' forms of matter (perfect 
fluids, minimally coupled scalar fields, Maxwell field, {\em 
etc.}) have a well-posed Cauchy problem, on the basis of the 
fact that they do in general 
relativity and that the relevant field equations 
do not change much in scalar-tensor gravity, once 
the gravitational sector is proved to be 
well-behaved with respect to the initial value 
problem \cite{Wald, HawkingEllis}.} and is well-posed 
in vacuo. It  was then straightforward to generalize  
this work to BD theories with constant BD parameter 
$\omega_0 \neq 1$ 
which, in turn, was used to show that the Cauchy problem of metric $f(R)$ 
gravity (equivalent to an $\omega_0=0$ BD theory) is 
well-formulated and well-posed in vacuo, while the Cauchy 
problem for Palatini $f(R)$ gravity (equivalent to an $\omega_0=-3/2$ BD 
theory)  is not 
well-formulated, nor 
well-posed \cite{LanahanFaraoni}. A second paper by Salgado and 
co-workers using a second order hyperbolicity analysis 
\cite{Salgadoetal} 
showed the well-posedness 
of $\omega=1$ theories, and the extension to $\omega=$const. 
theories (with the exception of $\omega=-3/2$) is 
straightforward because the principal part of the field 
equations does not depend on $\omega$.

In retrospect, it is easy to see  why the $\omega_0=-3/2$ BD 
theory 
does not admit a well-posed initial value formulation: the 
field equation for the  BD scalar is
\be \label{7}
\left( \omega_0+ \frac{3}{2} \right) \Box \phi 
+\frac{\omega_0}{2\phi} \, \nabla^c\phi \nabla_c\phi 
=\frac{T}{2\phi}+V'-\frac{2V}{\phi} \;,
\ee
and reduces to a first order constraint when $\omega_0\rightarrow 
-3/2$. Technically, this fact prevents the substitution of 
$\Box 
\phi$ back into the equations for the other dynamical variables 
in order to eliminate second derivatives of $\phi$ and spoils 
the reduction to a first-order system (see 
\cite{LanahanFaraoni} for details). In practice, the second 
order dynamical 
equation for the variable $\phi$ is lost when $\omega_0 =-3/2$, 
$\phi$ then plays the role of a non-dynamical auxiliary field 
and 
can be assigned 
arbitrarily {\em a priori}. Uniqueness of the solutions is then 
lost, as infinitely many prescriptions for $\phi$ correspond to 
the same set of initial data.

The present paper serves various purposes. First (Sec.~II), we 
revisit 
the Cauchy problem for BD theories and, in particular, for the 
equivalent of Palatini  $f(R)$ gravity by using  
a completely independent 
approach  based on the transformation to the Einstein conformal 
frame. 
This approach was deliberately avoided in previous papers 
\cite{Salgado, LanahanFaraoni, Salgadoetal}.  While 
interesting in itself as an independent  check of previous 
results, this approach has the additional merit of fully 
establishing 
the physical equivalence between Jordan and Einstein frames at 
the classical level. These conformal frames have been shown to 
be equivalent in 
various other respects, and it would only make sense that their 
equivalence extend to the Cauchy problem. 
However, there are explicit statements in the literature, 
and much unwritten folklore, pointing to the contrary. We show 
here that 
the 
two frames  are indeed  equivalent, which removes previous 
doubts and fully 
establishes equivalence at the classical level; however,  
this  does not guarantee physical equivalence at the  
quantum level \cite{Flanagan, FaraoniNadeau}.

The main purpose of this paper, however, consists of the study 
of the  Cauchy problem for scalar-tensor theories of the 
general  form~(\ref{3}) and of two distinct types of 
singularities that may appear in their field equations. These 
theories were not covered explicitly in 
previous literature, although the extension of the results of 
\cite{Salgado, Salgadoetal} to include them is relatively  
straightforward. In addition, it is handy to consider 
the general form~(\ref{3}) of the
 theory in order to specialize 
the results to any scalar-tensor theory simply by prescribing 
specific forms of the coupling functions $f(\phi)$ and 
$\omega(\phi)$ and of the potential $V(\phi)$. We approach the  
problem  in both the Jordan frame (Sec.~III) and the Einstein 
frame (Sec.~IV) obtaining, of course, the same results.

In general scalar-tensor theories, there are two 
kinds of singularities to deal with: 
those at which $f(\phi)=0$, and a second kind identified by 
$f_1(\phi) \equiv \omega( \phi) + 
\frac{3(f'(\phi))^2}{2f(\phi)}=0$, which generalizes the 
$\omega_0=-3/2$  pathology encountered in BD theory and in 
Palatini $f(R)$ gravity. Singularities of 
the first kind should normally be excluded by requiring that 
$f(\phi)>0$ for all values of $\phi$, and this requirement is 
sometimes made explicit in the general formalism ({\em 
e.g.}, \cite{FaresePolarski}); nevertheless, works 
incorporating these singularities recur often in the 
literature, especially in  cosmology.  

At the singularities of the second 
kind $f_1=0$ (which have been known for a long time in 
particular incarnations of scalar-tensor gravity 
\cite{FutamaseMaeda, FutamaseRothmanMatzner, Hosotani}),  a  
phenomenology similar to that of Palatini $f(R)$ gravity spoils 
the 
Cauchy problem for special forms of the coupling function 
$\omega(\phi)$, or for critical field values. While 
the scalar field is allowed to pass through these 
``singularities'' in an 
isotropic Friedmann-Lemaitre-Robertson-Walker (FLRW) universe 
\cite{Hosotani, FutamaseMaeda, FutamaseRothmanMatzner}, the 
points where $f_1(\phi)=0$ are known to give rise to curvature 
and shear 
singularities in the anisotropic case.  These singularities  
were discovered in the special case of nonminimally 
coupled scalar field cosmology (corresponding to 
$f(\phi)=\frac{1}{\kappa}- \xi \phi^2 $ and $\omega =1$) in 
the early universe \cite{Linde, Starobinsky, FutamaseMaeda, 
FutamaseRothmanMatzner} and also in black hole 
perturbations \cite{BronnikovKireev}. They also  appear in the search for 
exact wormhole 
solutions with  nonminimally coupled scalar fields 
\cite{BarceloVisser}. Singularities 
of both kinds  
were discussed in a more general context in \cite{AbramoBGS, 
VFsingularities}.  
After clarifying and further generalizing  
this situation from the point of view of the initial value 
formulation in Sec.~III, in Sec.~IV we 
revisit this subject in the Einstein frame, exposing a 
situation analogous to $\omega_0 =-3/2$ BD theory (this is not 
merely an analogy, since the latter is a special case of the 
former).  
Finally, in Sec.~V, we study nonminimally 
coupled scalar field 
theory as an example, recovering certain known properties and 
placing them in a general context. Sec.~VI and VII  
contain illustrative toy models and  the conclusions, respectively.

\section{\label{sec:II} II.~Einstein frame description of 
Brans-Dicke and Palatini $f(R)$ gravity}

In this section we recall the definition of Einstein 
conformal frame and show explicitly the non-dynamical role of 
the scalar field in the Einstein frame representation of the 
scalar-tensor version of 
Palatini $f(R)$ gravity. This is necessary as a first step to 
understand the more involved situation that we will be facing 
in later sections with general scalar-tensor theories of 
the 
form~(\ref{3}).

The conformal transformation
\be\label{8}
g_{ab}\rightarrow \tilde{g}_{ab}=\Omega^2 \, g_{ab}\;, 
\;\;\;\;\;\;\; \Omega=\sqrt{\phi} 
\ee
and the scalar field redefinition $\phi\rightarrow 
\tilde{\phi}$ with
\be\label{9}
d\tilde{\phi}=\sqrt{ \frac{ \left| 2\omega_0+3 \right|}{2\kappa} 
} \, \frac{d\phi}{\phi}
\ee
map the Jordan frame action of BD theory
\be\label{10}
S_{BD}=\int d^4x \, \sqrt{-g} \left[ \frac{\phi R}{2} 
-\frac{\omega_0}{2\phi} \nabla^c \phi \nabla_c \phi -V(\phi)   
+{\cal 
L}^{(m)} \right]
\ee
into its Einstein frame representation
\begin{eqnarray}
&& S_{BD} =  \int d^4x \, \sqrt{-\tilde{g}} \left[ 
\frac{ \tilde{R}}{2\kappa} 
-\frac{1}{2} \, \tilde{g}^{ab} \tilde{\nabla}_a \tilde{\phi} 
\tilde{\nabla}_b \tilde{\phi} -U( \tilde{\phi}) \right. 
\nonumber \\
&&\nonumber \\
&& \left. + \frac{ {\cal L}^{(m)} \left[ 
\Omega^{-1} \tilde{g}_{ab}, \psi 
\right] }{ \phi^2 }  \right] \;,
\label{11}
\end{eqnarray}
where 
\be\label{12}
U(\tilde{\phi})=\frac{ V( \phi ( \tilde{\phi})  
)}{\phi^2(\tilde{\phi})} 
\ee
and a tilde denotes rescaled (Einstein frame) quantities. 
The scalar field redefinition~(\ref{9}) 
breaks 
down when $\omega_0=-3/2$ and the scalar field $\tilde{\phi}$ 
then remains undefined. However, eq.~(\ref{8}) still holds and 
one can write the Einstein frame version of the BD 
equivalent of Palatini $f(R)$ gravity as \cite{review}
\begin{eqnarray}
S_{Palatini} &= & \frac{1}{2\kappa} \int d^4x \, 
\sqrt{-\tilde{g}} 
\left[ 
\tilde{R} - \frac{V(\phi)}{\phi^2} \right] 
\nonumber \\
&& \nonumber \\
&+& S^{(m)}\left[ 
\phi^{-1} \tilde{g}_{ab}, \psi \right] 
\label{13}
\end{eqnarray}
using the variables $\left( \tilde{g}_{ab}, \phi \right)$. 
In this action, the scalar field $\phi$ does not play any 
dynamical role: it only acts as a factor rescaling the metric 
in $S^{(m)}$ but it has no dynamics, does not couple to 
$\tilde{R}$,  and no kinetic energy of 
$\phi$  
appears in~(\ref{13}). $\phi$  can be assigned arbitrarily in 
infinitely many ways not governed by the usual second order 
differential equation and, therefore, uniqueness of the 
solutions is lost. On the contrary, for any value of the BD 
parameter $\omega_0 \neq -3/2$ (in particular for the $\omega_0=0$ 
equivalent of metric $f(R)$ gravity), the action is reduced 
to~(\ref{11}), which describes a scalar field $\tilde{\phi}$ 
coupling minimally to the curvature and nonminimally to 
matter. In vacuo (${\cal L}^{(m)}=0$), this coupling to matter disappears 
and we are left with the action of Einstein gravity plus a 
minimally coupled scalar  field with canonical kinetic energy: 
it is well-known that this system has a  well-posed initial 
value formulation \cite{Wald, HawkingEllis}. The non-vacuum 
case is considered later in Sec.~V as a special case   of 
more general scalar-tensor theories.

\section{\label{sec:III}III.~The Cauchy problem for general 
scalar-tensor theories in the Jordan frame}

Let us now restrict ourselves to the Jordan frame and consider 
general scalar-tensor theories described by 
the action~(\ref{3}). These can be reduced to the action
\be \label{questa}
S=\int d^4x\, \sqrt{-g}\left[ \frac{\Phi 
R}{2}-\frac{ \omega^* (\Phi)}{2}\, \nabla^c\Phi\nabla_c\Phi 
-U(\Phi)\right] +S^{(m)}
\ee
containing a  single coupling function $\omega^* (\Phi)$ by 
setting $\Phi \equiv f(\phi)$, $\omega^*(\Phi) = 
\omega(\phi(\Phi))$,  and $U(\Phi)=V(\phi(\Phi))$. The 
actions~(\ref{3}) and 
(\ref{questa}) are equivalent if $f(\phi)$ is invertible with 
regular inverse $f^{-1}$, but this does not happen if 
$f'(\phi)$ 
vanishes somewhere. 

It is also possible to recast BD theory as one in which the 
kinetic term of the scalar field is canonical, {\em 
i.e.}, with $ \omega=1$. 
Beginning with the action~(\ref{questa}) and setting
\be
\Phi = F(\varphi) \;,
\ee
where the function $F(\varphi)$ is defined by the equation
\be
\omega^* (\Phi) =\frac{F(\varphi)}{2\left( 
\frac{dF}{d\varphi} 
\right)^2} \;,
\ee
(\ref{questa}) can be rewritten as 
\begin{eqnarray}  
S&=&\int d^4x\, \sqrt{-g}\left[ \frac{F(\varphi )
R}{2}-\frac{1}{2}\, \nabla^c\varphi\nabla_c\varphi 
-W(\varphi)\right]  \nonumber \\
&&\nonumber \\
&+&  S^{(m)}\left[ g_{ab}, 
\psi \right] \;,
\label{nongeneralaction}
\end{eqnarray}
where $W(\varphi)=V \left[ F(\varphi )\right] $.
This alternative form of the BD action can not be obtained when 
$F(\varphi)$ does not admit a regular inverse $F^{-1}$ ({\em 
e.g.},  when $dF/d\varphi=0$). This is the case, for example, 
when $F(\varphi)$ is represented by a series of even powers of 
$\varphi$ \cite{LiddleWands, TorresVucetich}. Note 
that~(\ref{nongeneralaction}) is the form of the action 
considered in the studies of the Cauchy problem \cite{Salgado, 
Salgadoetal}. In what follows, to achieve full 
generality, we  discuss the action~(\ref{3}) with 
two coupling functions. Moreover, there are two types of 
``singularities'' to consider: we 
introduce  them here and we will refer to them for the rest of 
this  paper. In addition, one must distinguish betwen two very 
different situations: that in which these ``singularities'' 
occur in an entire four-dimensional domain of spacetime, and that in 
which they occur only on  
hypersurfaces. Morever, we will approach all of the 
above from the 
two viewpoints of Jordan frame and Einstein frame.

\subsection{Singularities of the first kind}

The first type of singularities is identified 
by $f(\phi_*)=0$ and occurs for critical values 
$\phi_*$ of the scalar field   
(if solutions of this equation exist). This 
equation may be satisfied in an entire four-dimensional spacetime 
region,  or 
on a hypersurface. At a first glance, the former case 
seems rather trivial: in fact, naively, the effective 
gravitational coupling read off the action~(\ref{17}) is  
$G_{eff}=1/f(\phi)$. However, a more careful analysis of the 
effective gravitational coupling in a Cavendish experiment, which is the 
only one directly accessible to local experiments, yields 
\cite{Nordvedt}
\be\label{20}
G_{eff}(\phi)= \frac{ 2\omega f +\left( 
2d f/d\phi  \right)^2 }{8\pi f \left[ 2\omega f + 3\left( 
df / d\phi \right)^2 \right] } \;.
\ee
This expression can also be obtained from the study of 
cosmological perturbations \cite{Boisseauetal}. The first type of 
singularities $f(\phi_*)=0$  corresponds to diverging effective 
coupling $ G_{eff}(\phi) $ 
and separates regions in which $G_{eff}$  has 
opposite signs describing attractive or repulsive 
gravity, respectively.  Stated this way, it may seem  
nonsensical to 
consider such values $\phi_* $ of the scalar. For example, it 
looks plain silly to consider, in BD teory, a spacetime region 
in which $\phi =0$, which makes the term $\phi R/2$ 
disappear 
from the BD action and corresponds to infinite strength of 
gravity.  Nevertheless, there are examples in which exact 
(and non-unique) solutions of the field equations have been found 
with $\phi$ constant and precisely equal to $\phi_*$ in a 
region, or in the entire spacetime manifold (see 
\cite{Hosotani, who?, 
superquintessence} for examples in cosmology and  
\cite{BarceloVisser} for wormhole solutions). Are these to be 
discarded {\em a priori}? Perhaps not, because what is 
clearly unphysical are regions in which $G_{eff}<0$ and 
the graviton is 
a ghost. Although rather pathological, regions in which 
$G_{eff}$ is divergent may still be interesting in exotic 
situations when the birth of the universe or the interior of a 
wormhole are  considered. Furthermore, these regions may still be 
relevant  from the mathematical point of view if one is 
interested in finding exact solutions that, as simplified toy models, 
exhibit  particular 
properties of scalar-tensor gravity.

Let us come now to the more interesting situation in which 
$f(\phi)=0$ {\em on an hypersurface}. This situation 
seems more 
reasonable, however such hypersurfaces  separate  regions 
of attractive from regions of repulsive 
gravity; in the latter, the graviton carries negative kinetic 
energy, a physically unacceptable property 
\cite{DamourFarese, FaresePolarski}. This fact 
seems to  be forgotten in scalar-tensor theories more general 
than BD theory and with   more freedom in the form of the 
functions $f(\phi)$ and $ 
\omega(\phi)$.  Papers in which $G_{eff} $ is negative 
or infinite   
have  appeared  surprisingly often  over the  past  thirty 
years \cite{Starobinsky, Hosotani, Linde, Gurevich, Pollock, 
GunzigNardone, Novello}; sometimes, such critical 
hypersurfaces $\phi_*$ are approached asymptotically  
\footnote{In \cite{who?}, a scenario  was proposed which exhibits  a 
singularity-free early universe 
with a  conformally coupled and self-coupled scalar 
field $\phi $ asymptotically emerging from a Minkowski space 
corresponding to the 
critical values $\phi_*$ 
in the past.}.

Let us proceed, for the moment, by adopting 
a purely mathematical point of view in the consideration of  
the Cauchy problem.  When $f(\phi)=0$,  
eq.~(\ref{4}) 
for the metric tensor degenerates. At these spacetime 
points  the trace of 
eq.~(\ref{4}) becomes 
\be\label{delta}
3\Box \phi +\left( \omega+3f''\right) 
\nabla^c\phi\nabla_c\phi +4V-T=0 \;.
\ee
Substitution of the value of $\Box\phi$ obtained from this 
equation into the second field equation~(\ref{5}) yields
\begin{eqnarray}
&& R=\frac{2}{f'}\left\{ \left[ \frac{\omega'}{2} 
-\frac{\omega}{3f'} 
\left( \omega +3f'' \right) \right] \nabla^c\phi \nabla_c\phi 
 \right.\nonumber \\
&&\nonumber \\
&&\left. +\frac{\omega}{3f'} \left( T-4V \right) -2V' \right\} \;.
\end{eqnarray}
Knowledge of the values of $\phi$ and of its gradient 
$\nabla_c\phi$ on the hypersurface $f=0$ determines the Ricci  
curvature. However, the equation for $ R_{ab}$ disappears 
there, which means that all metrics with the same value of 
$R$ satisfy the (degenerate) field equations on this 
hypersurface: uniqueness of the solutions is lost and this 
surface is a Cauchy horizon. The initial value problem breaks 
down at these critical points. Therefore, even if we 
decide to allow the unphysical region $G_{eff}<0$ by attempting to 
propagate 
initial data given in a $G_{eff}>0$ region, we encounter 
a hypersurface on which $G_{eff}\rightarrow \infty$ which acts as a 
barrier and 
the initial value formulation ceases to be well-posed.

\subsection{Singularities of the second kind}

Let us introduce now a second type of critical values 
of the scalar field that have previously been 
associated to physical  (curvature) singularities and 
that also correspond to a breakdown of the 
initial value problem.  Following the lesson of  
$\omega=-3/2$  theory \cite{LanahanFaraoni}, one 
notices that $\Box \phi $ disappears from the field 
equation~(\ref{5}) when  
\be\label{14}
f_1(\phi)\equiv \omega (\phi) +\frac{3(f'(\phi))^2}{2f(\phi) 
}=0 \;.
\ee
Again, one has to distinguish two cases: a)~eq.~(\ref{14}) is 
satisfied in a four-dimensional spacetime region, and b)~it is satisfied 
on a 
hypersurface. The former corresponds to regarding 
eq.~(\ref{14}) as specifying a particular form of 
the coupling function $\omega(\phi)$ (given $f(\phi)$), while 
the latter corresponds to seeing eq.~(\ref{14}) as a 
trascendental (or algebraic, depending on the forms of the 
functions $\omega$ and  $f$) equation 
that may admit as roots special critical values $\phi_c$ of the 
scalar   $\phi$  \footnote{{\em Cf.}  Ref.~\cite{Bronnikov} for 
conformal 
continuation past these points.}.

Let  us consider case~a) first: this is completely 
analogous to   
the  case of $\omega=-3/2$ BD theory which 
eq.~(\ref{14}) generalizes. When $f_1(\phi)$ vanishes 
identically for all values of $\phi$ in a four-dimensional  spacetime 
domain, the 
dynamics of the scalar $\phi$ are lost together with $\Box 
\phi$ and with the second order of the partial differential 
equation for $\phi$. The exception consists of situations in 
which the scalar satisfies $\Box \phi=0$, in which case there 
may be non-trivial dynamics for 
$\phi$, but this quantity disappears spontaneously from the 
field equations for the other variables. This situation 
includes general relativity with $\phi=$const. (for which the 
initial value problem is well-posed \cite{Wald} and the 
previous discussion obviously does not apply), and harmonic 
$\phi$-waves.

Situation~a) is, of course, the only possibility when $\omega$ 
represents a constant parameter instead of a function, as in BD 
theory. 
The general scalar-tensor theory is richer and 
allows one to contemplate the  possibility~b)  that 
eq.~(\ref{14}) is satisfied on a hypersurface. It is 
interesting that, in the absence of matter,  
invariants of the 
Riemann  tensor diverge at  this hypersurface for 
anisotropic metrics, while no such 
divergence occurs in isotropic 
FLRW spaces  
\cite{FutamaseRothmanMatzner, AbramoBGS}. Mathematically 
speaking, if $f(\phi)\neq 0$ and $f_1(\phi)$ is a 
continuous function, a hypersurface where 
$f_1(\phi)=0$ separates two regions corresponding to opposite 
signs of $f_1$ (unless the form of $f_1$ is pathologically fine-tuned): in 
each of these, the Cauchy problem may be 
well-posed but when one tries to propagate initial data through 
such a hypersurface, $\Box \phi$ given by 
\begin{eqnarray}   
 && \Box \phi =
\left\{ -\left( 
\frac{\omega '}{2} +\frac{3f'f''}{2f}+\frac{ \omega f'}{2f} 
\right) \nabla_c\phi \nabla^c \phi \right. \nonumber \\
&&\nonumber \\  
&& \left.+  \frac{f' T}{2f}\, 
  +2V'-\frac{2Vf'}{f} \right\} \left[ \omega+\frac{3(f')^2}{2f} \right]^{-1}
\end{eqnarray}
diverges. We have, therefore, a Cauchy horizon that is not 
hidden inside an apparent horizon, as in black holes, and where 
the theory crashes. The two regions separated by the 
hypersurface $f_1(\phi)=0$ are, again, disconnected by a 
singularity in the gravitational coupling $G_{eff}(\phi)$.

To summarize this section: when the coupling functions  
$f(\phi)$ and $\omega(\phi)$ are such that $f(\phi)=0$ or 
$f_1(\phi)=0$, the 
initial value formulation breaks down and either the 
theory is unphysical because $\phi$  becomes 
a non-dynamical auxiliary field, or the hypersurface 
$f_1(\phi)=0$ is a Cauchy horizon. In the first case, 
the 
problems found for Palatini $f(R)$ gravity  
in \cite{BarausseSotiriouMiller} re-surface. The situation 
in which eq.~(\ref{14}) is identically satisfied is the 
generalization to arbitrary scalar-tensor theories of the 
situation already seen in $\omega=-3/2$ BD theory and in  
Palatini $f(R)$ gravity.   The trace 
equation~(\ref{delta})  allows one to replace the trace $T$ with an 
expression containing second derivatives of $\phi$. Then,  the 
metric 
depends on derivatives of the scalar field of order higher than 
second and discontinuities, or irregularities, are not 
smoothed out by an integral of matter fields giving the metric $g_{ab}$ (for 
example, as in the usual 
Green function integral in the weak-field limit), but 
they cause  step-function discontinuities in the metric 
derivatives and  curvature singularities where the same matter 
distribution in Einstein's theory would generate a perfectly 
regular geometry.

\section{\label{sec:IV}IV.~General scalar-tensor theories and 
the 
Cauchy problem  in the Einstein frame}

We now examine the initial value problem of general 
scalar-tensor gravity in the Einstein frame. The conformal 
transformation
\be\label{15}
g_{ab}\rightarrow \tilde{g}_{ab}=\Omega^2 \, g_{ab}\;, 
\;\;\;\;\;\;\; \Omega=\sqrt{ f( \phi )} 
\ee
and the scalar field redefinition 
\be\label{16}
\tilde{\phi}=\int \sqrt{ \left| 2\omega f +3 (f')^2
\right| } \, \frac{d\phi}{f( \phi ) }
\ee
bring the Jordan frame action
\be\label{17}
S=\int d^4x \, \sqrt{-g} \left[ \frac{f(\phi) R}{2} 
-\frac{\omega (\phi) }{2} \nabla^c \phi \nabla_c \phi -V(\phi) 
+{\cal  L}^{(m)} \right]
\ee
into its Einstein frame representation 
\be\label{18}
S=\int d^4 x \, \sqrt{-\tilde{g}} \left[ \frac{ \tilde{R} 
}{2\kappa} 
-\frac{1}{2} \, \tilde{g}^{ab} \tilde{\nabla}_a 
\tilde{\phi} 
\tilde{\nabla}_b \tilde{\phi} -U( \tilde{\phi}) +
\frac{ {\cal L} ^{(m)} }{f^2 }  \right] \;,
\ee
where $ U(\tilde{\phi} ) = V( \phi( \tilde{\phi}) )/f^2$ and 
$f=f(\phi(\tilde{\phi}))$.
Again, apart from the now familiar coupling of the ``new'' 
scalar $\tilde{\phi}$ to matter described by ${\cal L}^{(m)} 
/f^2 $ (with the exception of 
conformally invariant matter), this action describes general 
relativity  with a canonical scalar field which couples 
minimally to the curvature but nonminimally to matter. As 
before, it is clear that the system has a well-posed initial 
value formulation in vacuo. This conclusion applies where the 
Einstein frame variables $\left( \tilde{g}_{ab} , \tilde{\phi} 
\right)$ are well-defined, {\em i.e.}, for $f(\phi)\neq 0$ and 
$f_1(\phi)\neq 0$.  It can be shown that the 
Cauchy problem is   well-posed in the presence of matter as 
well: this was already pointed out in 
ref.~\cite{FaresePolarski}, but is checked at the  end of this section  by 
extending the first order hyperbolicity 
analysis of \cite{Salgado}.

The  exception is when $f_1(\phi)=0$, 
in which case the scalar $\tilde{\phi}$ can not be defined 
using eq.~(\ref{16}). In this case, one can use the 
variables $\left( \tilde{g}_{ab}, \phi \right)$ instead of 
 $\left( \tilde{g}_{ab}, \tilde{\phi} \right)$,  
obtaining the Einstein frame action
\be\label{19}
S_{ST}=\int d^4x \, \sqrt{-\tilde{g} } \left[ \frac{\tilde{R}}{2\kappa}  
-\frac{ 
V(\phi) }{ f^2(\phi) } -\frac{ {\cal L}^{(m)}}{ f^2(\phi)} 
\right] \;.
\ee
Again, there are no dynamics for $\phi$ and the Cauchy problem 
is not well-formulated, nor well-posed, in this case due to the 
loss of uniqueness of the solutions. Morever, this result holds 
in both the Jordan and the Einstein frames, which then become 
physically equivalent in this respect.

The breakdown of the scalar field redefinition~(\ref{16}) is 
accompanied by other signals that something is going wrong with 
the physics when $f_1(\phi)=0$.  The effective 
gravitational coupling~(\ref{20}) diverges when $f_1=0$ (as a special 
case, $G_{eff}=\frac{2(2\omega_0+2)}{(2\omega_0+3) \phi} $ diverges 
as $\omega_0\rightarrow -3/2$ in BD theory). Moreover, it changes 
sign when $\phi$ crosses a critical value $\phi_*$ or $\phi_c$.
These critical values are attained by the scalar field 
in certain early universe inflationary scenarios with 
nonminimally coupled scalar fields, corresponding to 
$f(\phi)= \frac{1}{\kappa}-  \xi \phi^2 $ and $\omega=1$ ($\xi$ 
being a  dimensionless coupling constant) when $0<\xi<1/6$  
\cite{Starobinsky, 
FutamaseMaeda, FutamaseRothmanMatzner, Hosotani}. The same phenomenon 
in more general 
scalar-tensor theories is considered in \cite{AbramoBGS, 
VFsingularities}.

The authors of \cite{AbramoBGS} find that, 
in Bianchi cosmologies, the regions of the phase space 
at which $f_1(\phi)=0$ correspond to geometric singularities 
with divergent Kretschmann scalar $R_{abcd}R^{abcd}$. The 
$f_1=0$ singularity is  dynamically  forbidden in a closed or 
critically open  FLRW universe 
under the  assumptions $\rho \geq 0, V(\phi) \geq 0$, and 
$\omega \geq 1$  \cite{VFsingularities}.

The lesson of \cite{FutamaseRothmanMatzner, Starobinsky, AbramoBGS} is that, 
if there 
is even a small anisotropy, the 
change from attractive to repulsive gravity at $f_1=0$   can 
only occur through a  shear or curvature 
singularity which stops the evolution of the geometry: nature's 
message seems to be that  gravity can not spontaneously 
become repulsive in the absence of exotic matter violating the 
energy conditions (it is the purely gravitational sector of the theory 
that we are studyng here).

Note that in the theories considered by \cite{AbramoBGS, 
VFsingularities}, which have $\omega \equiv 1$, the singularity 
$f_1=0$ is automatically 
removed by requiring that $f(\phi) >0$ ({\em i.e.}, that 
the graviton is 
not a ghost); however, this is no longer true when theories 
with $\omega$ not identically equal to unity are considered and 
critical values $\phi_c$ of the second kind can still occur 
even when $f(\phi)>0$ $\,\forall \, \phi$ --- but this necessarily 
requires 
$\omega<0$.

Let us now extend the first order hyperbolicity 
analysis of \cite{Salgado} to Einstein frame scalar-tensor gravity. We follow 
closely, and adopt the notations of,   
\cite{Salgado, LanahanFaraoni} in order to facilitate comparison, setting  
$\kappa = 1$.  The Einstein frame field equations are
\begin{eqnarray}
 \tilde{G}_{ab} & =&    
\tilde{\nabla}_{a} \tilde{\phi} \tilde{\nabla}_{b}\tilde{\phi} - \frac{1}{
2} \tilde{g}_{ab} \, \tilde{\nabla}^{c} \tilde{\phi} 
\tilde{\nabla}_{c}\tilde{\phi}  - U(\tilde{\phi}) \tilde{g}_{ab} \nonumber 
\\
&&\nonumber \\
& +& 
\frac{T_{ab}^{(m)}}{f^2(\phi(\tilde{\phi}))} \equiv 
\tilde{T}_{ab}[\tilde{\phi}] 
+\tilde{T}^{(m)}_{ab}\equiv \tilde{T}_{ab}  \;,\label{A24}\\
&&\nonumber \\
&& \tilde{\square}\tilde{\phi} -  
\frac{d U ( \tilde{\phi})}{d\tilde{\phi} } = 0 \;.
\end{eqnarray}
Because $T_{ab}[\tilde{\phi}] $ does not contain second derivatives of 
$\tilde{\phi}$, it is possible to give a first order formulation as in 
general relativity. The  nonminimal coupling factor  
$1/f^2(\phi(\tilde{\phi}) )$ multiplying $T_{ab}^{(m)}$ on the right 
hand side of eq.~(\ref{A24})  
does not generate derivatives of $\tilde{\phi}$ and therefore is immaterial.

The $3+1$ ADM formulation of the theory  defines the  
usual lapse, shift,  extrinsic curvature, and 
gradients of $\phi$ 
\cite{Wald,  Salgado}. Assuming  the existence of  a time function $t$ 
such that the spacetime ($M$,$ \tilde{g}_{ab}$) is foliated by a family 
of  hypersurfaces $\Sigma_{t}$ of constant $t$ with unit timelike normal 
$ \tilde{n}^{a}$, the 
3-metric is defined by $ \tilde{h}_{ab} = \tilde{g}_{ab}+\tilde{n}_{a}
\tilde{n}_{b}$ 
and 
${\tilde{h}^{a}}_{c}$  is the projection operator on $\Sigma_{t}$. The 
relations 
$ \tilde{n}^{a} \tilde{n}_{a} = -1 $, $ 
\tilde{h}_{ab} \tilde{n}^{b} = \tilde{h}_{ab
}\tilde{n}^{a} = 0 $, and $  {\tilde{h}_{a}}^{b} 
\tilde{h}_{bc} = 
\tilde{h}_{ac} $ 
are satisfied. Further introducing the 
lapse $\tilde{N}$, shift vector $ \tilde{N}^{a}$, and spatial metric 
$\tilde{h}_{ij}$, the metric is written  as
\begin{equation}
d\tilde{s}^{2} = -\left( \tilde{N}^{2} - \tilde{N}^{i} \tilde{N}_{i} \right) 
dt^{2} - 2\tilde{N}_{i}dtdx^{i} + \tilde{h}_{ij}dx^{i}dx^{j}
\end{equation}
$\left( i,j = 1,2,3 \right)$, with $\tilde{N} > 0$, $\tilde{n}_{a} = - 
\tilde{N} \tilde{\nabla}_{a} t
$ and 
\begin{equation}
\tilde{N}^{a} = -{\tilde{h}^{a}}_{b}t^{b} \;,
\end{equation}
where the time flow vector $ \tilde{t}^{a}$ satisfies $ 
\tilde{t}^{a}  \tilde{\nabla}_{a}t = 1$ and
\begin{equation}
\tilde{t}^{a} = -\tilde{N}^{a}+\tilde{N} \tilde{n}^{a} 
\end{equation}
so that $ \tilde{N} = - \tilde{n}_{a} \tilde{t}^{a}$ and $ \tilde{N}^{a} 
\tilde{n}_{a} = 0$. The extrinsic curvature 
 of $\Sigma_{t}$ is
\begin{equation}
\tilde{K}_{ab} = -{\tilde{h}_{a}}^{c} 
{\tilde{h}_{b}}^{d}\tilde{\nabla}_{c} \tilde{n}_{d} \;. 
\end{equation}
The 3D covariant derivative of $\tilde{h}_{ab}$ on $\Sigma_{t}$ is 
defined as
\begin{equation}
\tilde{D}_{i} ^{\left(3\right)}{ T^{a_{1}\ldots}}_{b_{1}\ldots} = { 
\tilde{h}^{a_{1}}  }_{c_{1}}  \ldots 
{\tilde{h}^{d_{1}}}_{b_{1}}\ldots {\tilde{h}^{f}}_{i}\tilde{\nabla}_{ f} ^{ 
\left(3\right)} {T^{c_{1}\ldots}}_{d_{1}\ldots}
\end{equation}
for any 3-tensor $^{(3)} {T^{a_{1} \ldots} }_{b_{1}\ldots}$ ,  with $ 
\tilde{D}_i \tilde{h}_{ab}= 0$.  The  spatial gradient of the scalar field 
and its momentum are 
\begin{equation}
\tilde{Q}_a \equiv  \tilde{D}_a \tilde{\phi} \;,
\end{equation}
and
\begin{equation}
\tilde{\Pi} = {\cal L}_{\tilde{n}} \tilde{\phi} = 
\tilde{n}^c \tilde{\nabla}_c \tilde{\phi} \;,
\end{equation}
respectively, and 
\begin{equation}
\tilde{K}_{ij} = -\tilde{\nabla}_i \tilde{n}_j = 
-\frac{1}{2\tilde{N}}\left(\frac{\partial \tilde{h}_{ij}}{\partial t}  + 
\tilde{D}_i \tilde{N}_j 
+ \tilde{D}_j \tilde{N}_i \right) \;,
\end{equation}
\begin{equation} 
\tilde{\Pi} = \frac{1}{\tilde{N}}\left(\partial_t 
\tilde{\phi}+\tilde{N}^c \tilde{Q}_c\right) \;,
\end{equation}
\begin{equation}
\partial_t \tilde{Q}_i+ 
\tilde{N}^l \partial_l \tilde{Q}_i+ 
\tilde{Q}_l\partial_i \tilde{N}^l= \tilde{D}_i\left( \tilde{N} 
\tilde{\Pi} \right) \;. 
\end{equation}
The stress-energy tensor is $3+1$-decomposed as 
\begin{equation}
\tilde{T}_{ab} = \tilde{S}_{ab}+ \tilde{J}_a \tilde{n}_b+ \tilde{J}_b 
\tilde{n}_a+ \tilde{E} \tilde{n}_a \tilde{n}_b  \;,
\end{equation}
where
\begin{equation} \label{A41}
\tilde{S}_{ab} \equiv {\tilde{h}_a}^c \, {\tilde{h}_b}^d \tilde{T}_{cd} 
= 
\tilde{S}_{ab} [\tilde{\phi}]  
+\tilde{S}_{ab}^{(m)} \;,
\end{equation}
\begin{equation}
\label{A42}
\tilde{J}_{a} \equiv -{\tilde{h}_a}^c \, \tilde{T}_{cd} \, \tilde{n}^d = 
\tilde{J}_{a} [\tilde{\phi}] + \tilde{J}_{a}^{(m)} \;,
\end{equation}
\begin{equation}\label{A43}
\tilde{E} \equiv \tilde{n}^a \tilde{n}^b \tilde{T}_{ab} = 
\tilde{E}[ \tilde{\phi} ] +\tilde{E}^{(m)} \;,
\end{equation}
and $\tilde{T} = \tilde{S}-\tilde{E}$, where $\tilde{T} $ is the trace of 
$\tilde{T}_{ab}$  and $\tilde{S}$  is the trace of $\tilde{S}_{ab}$.
The Gauss-Codacci equations provide the  Einstein equations 
 projected tangentially and orthogonally to $\Sigma_t$ as the Hamiltonian 
constraint  \cite{Wald, Salgado}
\begin{equation}
^{(3)}\tilde{R} + \tilde{K}^2 - \tilde{K}_{ij} \tilde{K}^{ij} = 2\tilde{E} 
\;,
\end{equation}
the vector (or momentum) constraint 
\begin{equation}
\label{A45}
\tilde{D}_l {\tilde{K}^l}_i - \tilde{D}_i \tilde{K} = \tilde{J}_i \;,
\end{equation}
and the dynamical equations
\begin{eqnarray}
\label{A46}
&& \partial_t {\tilde{K}^i}_j + \tilde{N}^l \partial_l {\tilde{K}^i}_j + 
{\tilde{K}^i}_l \partial_j \tilde{N}^l - 
{\tilde{K}^l}_j 
\partial_l \tilde{N}^i + \tilde{D}^i \tilde{D}_j \tilde{N} \nonumber \\
&&\nonumber \\
&&- ^{(3)}{\tilde{R}^i}_j \tilde{N} - \tilde{N} \tilde{K}{\tilde{K}^i}_j = 
\frac{\tilde{N}}{2}\left[ \left(\tilde{S}-\tilde{E}\right) 
\delta^i_j -2 \tilde{S}^i_j 
\right] ,
\end{eqnarray}
where $\tilde{K} \equiv {\tilde{K}^i}_i$. The trace of this equation yields
\begin{equation}
\partial_t \tilde{K} + \tilde{N}^l \partial_l \tilde{K} + 
^{(3)}\tilde{\Delta} \tilde{N} - \tilde{N} \tilde{K}_{ij} \tilde{K}^{ij} = 
\frac{\tilde{N}}{2} \left( \tilde{S} + \tilde{E} \right) \;,
\end{equation}
where $^{(3)}\tilde{\Delta} \equiv \tilde{D}^i \tilde{D}_i$.

Further introducing  $\tilde{Q}^2 \equiv \tilde{Q}^c \tilde{Q}_c$, one 
computes
\begin{equation}
\tilde{E} [ \tilde{\phi}]  = \frac{1}{2}\left( \tilde{\Pi}^2 + \tilde{Q}^2 
\right) + U( \tilde{\phi}) \;,
\end{equation}
\begin{equation}
\tilde{J} [ \tilde{\phi}] = -\tilde{\Pi} \tilde{Q}_a \;,
\end{equation}
\begin{equation}
\tilde{S}_{ab} [ \tilde{\phi}] = \tilde{Q}_a \tilde{Q}_b - 
\tilde{h}_{ab}\left[ \frac{1}{2}\left( 
\tilde{Q}^2 - \tilde{\Pi}^2 \right) + U( \tilde{\phi}) \right] \;,
\end{equation}
while
\begin{equation}
\tilde{S} [\tilde{\phi}] = \frac{a}{2}\left( 3\tilde{\Pi}^2 - \tilde{Q}^2 
\right) - 3U (\tilde{\phi}) 
\end{equation}
and
\begin{equation}
\tilde{S}[ \tilde{\phi}] -\tilde{E} [\tilde{\phi}] = \left( \tilde{\Pi}^2 - 
\tilde{Q}^2 \right) - 
4U (\tilde{\phi}) \;.
\end{equation}
The ``total'' quantities entering the right hand side  of the $3+1$ 
field 
equations are then
\begin{equation}
\tilde{E} =    
   \frac{1}{2} \tilde{Q}^2 + 
\frac{1}{2}\tilde{\Pi}^2 + U(\tilde{\phi}) + 
 \tilde{E}^{(m)}  \;,
\end{equation}

\begin{equation}
\tilde{J}_a =  
-   \tilde{\Pi} \tilde{Q}_a + \tilde{J}_a^{(m)}  \;,
\end{equation}

\begin{eqnarray}
\tilde{S}_{ab} &=&   
 -\tilde{h}_{ab}\left[  \frac{1}{2}  
\left( \tilde{Q}^2 - 
\tilde{\Pi}^2 \right) + U(\tilde{\phi})  \right] 
\nonumber \\
&&\nonumber \\
& +& \tilde{Q}_a \tilde{Q}_b + \tilde{S}_{ 
ab}^{(m)}  \;,
\end{eqnarray}
while
\begin{eqnarray} 
&& \tilde{S} =  - 3U(\tilde{\phi}) -    \frac{\tilde{Q}^2}{2}
 -   \frac{3\tilde{\Pi}^2}{2}  + \tilde{S}^{(m)} \;,\\
&&\nonumber\\
&& \tilde{S} - \tilde{E} =  \nonumber \\ 
&&\nonumber\\
&&   
 \tilde{\Pi}^2-\tilde{Q}^2 - 4U(\tilde{\phi}) + \tilde{S}^{(m)} 
- \tilde{E}^{(m
)}  \;,\\
&&\nonumber\\
&& \tilde{S} + \tilde{E} = 
2 \tilde{\Pi}^2  - 2U( \tilde{\phi})  + 
\tilde{S}^{(m)}+ \tilde{E}^{(m)} 
\;.
\end{eqnarray}
The Hamiltonian constraint becomes
\begin{eqnarray}
\label{A64}
&& ^{(3)}\tilde{R} + \tilde{K}^2 - \tilde{K}_{ij} \tilde{K}^{ij} 
 + \frac{\tilde{\Pi}^2}{2} + \frac{\tilde{Q}^2}{2}  \nonumber \\
&&\nonumber \\
&&= 
 \tilde{E}^{(m)}+U(\tilde{\phi})  \;,\nonumber \\
&& 
\end{eqnarray}
while the momentum constraint (\ref{A45}) is 
\begin{equation} 
\label{A65} 
\tilde{D}_l {\tilde{K}^l}_i - \tilde{D}_i \tilde{K} 
 +  \tilde{\Pi} \tilde{Q}_i  = 
\tilde{J}_i^{(m)} \;, 
\end{equation}
the dynamical equation~(\ref{A46}) is written as
\begin{eqnarray}
&& \partial_t {\tilde{K}^i}_j + \tilde{N}^l \partial_l {\tilde{K}^i}_j + 
{\tilde{K}^i}_l 
\partial_j \tilde{N}^l - {\tilde{K}_j}^l  \partial_l \tilde{N}^i 
+ \tilde{D}^i \tilde{D}_j \tilde{N}  \nonumber \\
&&\nonumber \\
&& - ^{(3)}{\tilde{R}^i}_j \tilde{N} - 
\tilde{N} \tilde{K}{ \tilde{K}^i}_j + \frac{\tilde{N}}{2}   2U(\tilde{\phi}) 
\delta^i_j  + \tilde{N} \tilde{Q}^i \tilde{Q}_j \nonumber \\
&&\nonumber \\
&& 
= \frac{\tilde{N}}{2} \left[\left( \tilde{S}^{(m)} - \tilde{E}^{(m)} \right) 
\delta^i_j - 2{\tilde{S}^{(m) \,\, i}}_j \right] 
\end{eqnarray}
with trace
\begin{eqnarray}
\label{A67}
&& \partial_t \tilde{K} + \tilde{N}^l \partial_l \tilde{K} + 
^{(3)}\tilde{\Delta} \tilde{N} - \tilde{N} \tilde{K}_{ij} \tilde{K}^{ij} - 
 \nonumber \\
&&\nonumber \\
&& -\tilde{N} \tilde{\Pi}^2  = 
\frac{\tilde{N}}{2}\left( -2U(\tilde{\phi})  
 + 
\tilde{S}^{(m)} + 
\tilde{E}^{(m)} \right) 
\end{eqnarray}
where \cite{Salgado}
\begin{eqnarray}
\label{A68}
&& {\cal L}_{\tilde{n}} \tilde{\Pi} - \tilde{\Pi} \tilde{K} - \tilde{Q}^c 
\tilde{D}_c \left( \ln \tilde{N} \right) - \tilde{D}_c 
\tilde{Q}^c  = -\tilde{\square}\tilde{\phi}
 \nonumber \\
&& =  -   \frac{dU}{d\tilde{\phi}}  \;.
\end{eqnarray}
In vacuo, the  initial data $\left (\tilde{h}_{ij}, \tilde{K}_{ij}, 
\tilde{\phi}, 
\tilde{Q}_i, \tilde{\Pi} \right)$ 
on an initial  hypersurface $\Sigma_0$ obey the  
constraints~(\ref{A64}) and 
(\ref{A65}) plus 
\begin{equation}
\tilde{Q}_i - \tilde{D}_i \tilde{\phi} = 0 \;, \;\;\;\;\;\;\;\;\;\;
\tilde{D}_i \tilde{Q}_j = \tilde{D}_j \tilde{Q}_i \;. 
\end{equation}
 In the presence of matter, the variables $\tilde{E}^{(m)}$, 
$\tilde{J}_a^{(m)}$, and $\tilde{S}_{ab}^{(m)}$ are also assigned
on the  initial hypersurface. Fixing a gauge corresponds to  
prescribing lapse  and shift. The system (\ref{A64})-(\ref{A67}) 
contains only  
first-order  derivatives 
in both space and time once  the d'Alembertian $\tilde{\square} 
\tilde{\phi}$ 
is written in terms of 
$\tilde{\phi}, \tilde{\nabla}^c \tilde{\phi} \tilde{\nabla}_c 
\tilde{\phi}$,  
and their derivatives by using eq.~(\ref{A68}). 
From this point on, 
everything proceeds as in 
 Ref.~\cite{Salgado} and the nonminimal coupling factor $ 
f(\phi(\tilde{\phi}))$ in $\tilde{T}_{ab}^{(m)}=T_{ab}^{(m)}/f^2$ 
does not have consequences because it contains no derivatives of 
$\tilde{S}_{ab}^{(m)}, \tilde{J}^{(m)}_a$, or $\tilde{E}^{(m)}$. The 
reduction 
to a first-order system  indicates that the Cauchy problem is 
well-posed in vacuo and well-formulated 
 in the presence of those forms of matter for which it is 
well-formulated in general relativity.  We do not duplicate Salgado's 
analysis  here, and we refer the 
reader to \cite{Salgado, Salgadoetal} for details.

\subsection{\label{sec:frames}Equivalence between conformal 
frames}

At this point, it is clear  that the 
initial 
value 
formulation is well-posed in the 
Einstein frame if it is well-posed in the Jordan frame, and 
vice-versa. The two frames are equivalent also from the 
point of view of the Cauchy problem, contrary to folklore and 
recurring statements in the literature.  To this regard, 
it is often remarked that the mixing of the spin two and 
spin zero degrees  of freedom $g_{ab}$ and $\phi$ in the Jordan 
frame makes these  variables an inconvenient set for 
formulating the initial value  problem, which is 
consequently not well-posed in the Jordan frame, while  the 
Einstein frame variables 
$\left(\tilde{g}_{ab},\tilde{\phi}\right)$ admit a well-posed 
Cauchy problem completely similar to that of general relativity.  
(A rather casual remark in  the well-known paper  
\cite{DamourFarese} (see also the more 
recent Ref.~\cite{FaresePolarski}) 
seems to have been quite influential in 
this respect, without further questioning of it in later  literature 
until the recent  work of Salgado \cite{Salgado}). In the 
light of this work, which is carried out completely in 
the Jordan frame, the standard lore is obviously false. 
Old works also hinted to the fact  that the  Cauchy 
problem is  
well-posed 
{\em in the Jordan frame} for  two  special scalar-tensor 
theories: Brans-Dicke theory with a free scalar $\phi$ 
\cite{CockeCohen}, and the 
theory of a scalar field conformally coupled to the Ricci 
curvature \cite{Noakes}.  The implementation, in the Jordan 
frame,  of a full 3+1  formulation $\grave{a}$ la 
York  \cite{York} for use in  numerical 
applications further dispels the myth that the Cauchy 
problem is not well-posed in the Jordan frame \cite{Salgadoetal}.

Were this folklore true,  the Jordan and Einstein 
frames would be physically  inequivalent with regard to the 
Cauchy problem, but we have shown that this is not the case. In 
fact,  the equivalence 
between the two conformal  frames does not break down even 
when the scalar field redefinition $\phi \rightarrow 
\tilde{\phi} $ fails.  
The Jordan and Einstein frame are still equivalent, with respect 
to the initial value formulation, for general scalar-tensor 
theories and, therefore, they are equivalent at the classical 
level, thus dissipating residual doubts left in this regard in 
\cite{FaraoniNadeau}.  However, the two conformal frames seem 
to 
be inequivalent at the quantum level  (\cite{Flanagan, 
FaraoniNadeau}  and references therein).

\section{\label{sec:V}V.~Example: the non-minimally coupled 
scalar 
field}

We are finally ready to consider, as an example, the theory of 
a scalar field coupled nonminimally  to the Ricci 
curvature. In 
fact this example,  many features of which are well-known, has 
sometimes already guided us through this paper. The action is 
\begin{eqnarray}
&& S_{NMC} = \int d^4 x \, \sqrt{-g} \left[ \left( 
\frac{1}{2\kappa}-\frac{\xi \phi^2}{2} \right) R -\frac{1}{2}\, 
\nabla^c\phi\nabla_c\phi \right.  \nonumber \\
&& \left. \right. \nonumber \\
& & \left. -   V(\phi)  +  \alpha_m {\cal L}^{(m)} \right] \;, 
\label{NMCaction}
\end{eqnarray}
where $\xi$ is a dimensionless coupling constant (in our 
notations, conformal coupling corresponds to $\xi=1/6$), and $\alpha_m$ 
is  a suitable coupling 
constant.  The field equations are
\begin{widetext}
\begin{eqnarray}
&& \left( 1-\kappa \xi \phi^2 \right) G_{ab}= 
\kappa \left[ 
\nabla_a\phi\nabla_b\phi -\frac{1}{2} 
g_{ab} \nabla^c\phi\nabla_c\phi -V(\phi) g_{ab} +\xi \left( 
g_{ab}\Box -\nabla_a\nabla_b \right) \left( \phi^2 \right) 
+T_{ab}^{(m)} \right] \;,  \label{NMC1} \\
&& \Box \phi-\frac{dV}{d\phi}-\xi R \phi =0  \label{NMC2}
\end{eqnarray}
\end{widetext}
(see \cite{BellucciFaraoni, mybook} for a discussion of alternative 
ways of writing the field equations).
By neglecting the matter part of the action, the Ricci curvature can be 
eliminated from the Klein-Gordon equation obtaining
\begin{eqnarray}
&& \frac{ 1+ \left( 6\xi -1 \right) \kappa\xi \phi^2}{1-\kappa\xi \phi^2} \, 
\Box \phi -\frac{\kappa \xi \phi}{1-\kappa \xi \phi^2} \left[ \left( 1-6\xi 
\right) \nabla^c\phi\nabla_c\phi \right. \nonumber \\
&&\nonumber \\
&& \left. +4V \right] -\frac{dV}{d\phi}=0  \;. \label{KGwithRsubstituted}
\end{eqnarray}
Singularities of the first kind correspond to $f(\phi)=\frac{1}{\kappa}-\xi 
\phi^2=0$, or to the  critical scalar field values
\be
\phi= \pm \phi_* \equiv \frac{\pm 1}{\sqrt{\kappa \xi}}
\ee
and can only occur if $\xi>0$. They correspond to diverging effective 
gravitational coupling
\be
G_{eff}=\frac{G}{1-\kappa\xi \phi^2} \;, 
\ee
which changes sign if the scalar $\phi$ crosses $\pm 
\phi_*$. The  requirement $f(\phi)>0 \; \forall \phi$ avoids these critical 
values. However, one could decide to momentarily ignore 
the physical interpretation of the theory and to allow 
these critical values from a purely mathematical point of view; 
then, the latter return to haunt  
the Cauchy problem and predictability.

The quantity $f_1$ is, in this theory,
\be
f_1(\phi)=\frac{1+\kappa\xi\left( 6\xi -1 \right) 
\phi^2}{1-\kappa\xi\phi^2} \;.
\ee
The roots of the equation $f_1=0$, which exist if 
 $0< \xi < 1/6$, are the critical values of 
the  second kind 
\be
 \pm \phi_c \equiv  \frac{\pm 1}{ \sqrt{ 
\kappa\xi\left(1-6\xi\right)}} \;.
\ee
The non-uniqueness of the 
solutions and the breakdown of the Cauchy problem marked by the 
critical values $\pm \phi_*, \pm \phi_c$ are seen as follows. 
When $\phi=\phi_0=$const. and matter is absent, the theory 
reduces to vacuum general relativity with a 
cosmological constant, and the field equations reduce to  
\begin{eqnarray}
&& G_{ab}+  \Lambda g_{ab}=0 \;, \;\;\;\;\; \Lambda=\frac{\kappa 
V(\phi_0)}{1+\kappa\xi\phi_0^2} \;, \label{ciccio1}\\
&&\nonumber \\
&& V'_0 +\xi R \phi_0=0 \;. \label{qquesta}
\end{eqnarray}
The trace of eq.~(\ref{ciccio1}) gives $R=4\Lambda$ which, compared 
with eq.~(\ref{qquesta}) in turn implies that 
\be
R=\frac{- V_0'}{\xi \phi_0}  \;.
\ee
If $\phi=\pm \phi_*$ (the critical values of the first kind), then it must be 
$V_0=0$ 
and, therefore $R_{ab}=0$. The Klein-Gordon equation 
yields the extra necessary condition $V_0'=0$. All vacuum solutions 
of general relativity ($R_{ab}=0$)  are also 
solutions of the field equations~(\ref{NMC1}) and 
(\ref{NMC2}) with $\phi=\pm \phi_*$.

If instead $\phi=\pm \phi_c $ (the second kind of critical values),  
eqs.~(\ref{ciccio1}) and (\ref{qquesta}) yield
\be
\Lambda=\frac{\kappa\left(1-6\xi \right) \, V(\pm\phi_c)}{2(1-3\xi)} \;,
\ee
and 
\be
V'( \pm \phi_c)=\mp \, \frac{2\sqrt{ \kappa\xi(1-6\xi)} \, V(\pm 
\phi_c)}{1-3\xi} \;.
\ee
At these critical scalar field values of the second kind, the dynamical 
equation~(\ref{KGwithRsubstituted}) for $\phi $ loses all the second 
derivatives of $\phi$ (contained in $\Box \phi$) and, consequently, the 
dynamics for this field (except for special solutions satisfying  
$\Box \phi=0$). In  isotropic FLRW spaces, 
solutions are known which cross the critical values $\pm \phi_c$, or $\phi$ 
is identically equal to one of these values. However, the  situation can 
be worse: there are physical curvature and shear singularities in anisotropic 
Bianchi  models \cite{Starobinsky, 
FutamaseRothmanMatzner, AbramoBGS,GunzigSaa}. Moreover,  Barcelo 
and  Visser \cite{BarceloVisser} find diverging Ricci scalar $R$ for 
spherically symmetric wormhole solutions. These examples  correspond to 
solutions which  cannot cross the barrier $\phi=\pm \phi_c$.

The conformal transformation to the Einstein frame is $g_{ab}\rightarrow 
\tilde{g}_{ab}=\Omega^2 \, g_{ab}$ with
\be
\Omega=\sqrt{1-\kappa \xi\phi^2}
\ee
and the redefinition bringing the scalar field into canonical form is
\be
d\tilde{\phi}=\frac{ \sqrt{ 1-\kappa \xi \left(1-6\xi 
\right)}}{1-\kappa\xi\phi^2} \, d\phi \;.
\ee
By integrating the last equation, the Einstein frame  scalar  $\tilde{\phi}$ 
can be explicitly expressed in terms of $\phi$ as 
\be  \label{nmcphif}
\tilde{\phi}= \sqrt{ \frac{3}{2\kappa}} \ln \left[ 
\frac{ \xi \sqrt{6\kappa\phi^2} + \sqrt{1-\xi \left( 1-6\xi \right) \kappa
\phi^2}}
{\xi \sqrt{6\kappa\phi^2} -\sqrt{1-\xi \left( 1-6\xi \right) \kappa
\phi^2}}  \, \right] + f \left( \phi \right) \; ,
\ee
where 
\be 
f \left( \phi \right) = 
\left( \frac{1-6\xi}{\kappa \xi} \right)^{1/2} \arcsin
\left( \sqrt{ \xi \left( 1-6\xi \right)  \kappa\phi^2} \right)
\ee
for $0 < \xi < 1/6 $ and
\be  \label{nmcphi2}
f \left( \phi \right) = 
\left( \frac{6\xi -1 }{\kappa \xi}  \right)^{1/2} 
\mbox{ arcsinh}
\left( \sqrt{ \xi \left( 6\xi -1 \right)  \kappa \phi^2}  \, \right) 
\ee
for $ \xi > 1/6 $, while 
\be  
\tilde{\phi}= \sqrt{ \frac{3}{2\kappa}} \ln  \left(  \frac{ 
 \sqrt{ 6/ \kappa} + \phi }{
 \sqrt{ 6/ \kappa} - \phi }  \right) \;\;\;\;\;\;\;\;\;\;
\mbox{if} \;\; \left| \phi \right| < \sqrt{ \frac{6}{\kappa} } \;,
\ee
or
\be  
\tilde{\phi}= \sqrt{ \frac{3}{2\kappa}} \ln  \left(  \frac{ 
 \phi - \sqrt{ 6/ \kappa} }{
 \phi+  \sqrt{ 6/ \kappa}  }  \right) \;\;\;\;\;\;\;\;\;\;
\mbox{if} \;\; \left| \phi \right| > \sqrt{ \frac{6}{\kappa} } 
\ee
for  $\xi=1/6$. 

The Einstein frame action is 
\begin{eqnarray}
S&=& \int d^4 x \, \sqrt{-\tilde{g}} \, \left\{ 
\frac{\tilde{R}}{2\kappa}
 -\frac{1}{2} \, \tilde{g}^{ab} \, \tilde{\nabla}_a \tilde{\phi}
\, \tilde{\nabla}_b \tilde{\phi} -U \left( \tilde{\phi} \right) 
\right. \nonumber \\
&&\nonumber \\
&& \left. +\tilde{\alpha}_m \left( \phi \right)  \, {\cal L}^{(m)} \right\}
\end{eqnarray}
where
\be \label{ciuccio}
U \left( \tilde{\phi} \right)= \frac{ V\left[ \phi ( \tilde{\phi} )
\right]}{\left[ 1-\kappa\xi \phi^2(\tilde{\phi}) \right]^2} 
\ee
and
\be
\tilde{\alpha}_m \left( \tilde{\phi} \right)= \frac{ \alpha_m }{ \left[
1-\kappa\xi \phi^2 ( \tilde{\phi}) \right]^2 } \;.
\ee
When  $\phi=\pm \phi_*$, the conformal transformation of the metric breaks 
down, 
while the redefinition of the scalar  field becomes invalid when 
$\phi=\pm \phi_c$. 
In this last situation, one can still use the variables 
$\left( \tilde{g}_{ab}, \phi \right)$ 
to define an Einstein frame in which the action is simply
\be
S=\int d^4x \sqrt{-\tilde{g}} \, \left( 
\frac{\tilde{R}}{2\kappa}-\frac{V(\phi)}{\left( 1-\kappa\xi 
\phi^2\right)^2}+\frac{\alpha_m}{\left( 1-\kappa\xi \phi^2\right)^2} \, {\cal 
L}^{(m)} \right)
\ee
with no dynamics for $\phi$, which  becomes an auxiliary field and can 
be  
assigned 
arbitrarily.

\section{\label{sec:VI}VI.~Toy models}

In this section we consider toy models in the context of point particle 
dynamics, which help obtaining some insight into the ``singularities'' of the 
first and second kind of scalar-tensor theories.

Let us first consider the point particle action
\begin{eqnarray} 
S&= & \int dt\, L\left( x(t), \dot{x}(t),y(t), \dot{y}(t) \right) 
\nonumber \\
&&\nonumber \\
&  = & \int dt \, 
\left[ \frac{\dot{x}^2 f(y)}{2}-\frac{w(y)\dot{y}^2}{2} -J(x)\right] \,
\label{toymodelaction}
\end{eqnarray}
where an overdot denote differentiation with respect to  
 the time $t$, the generalized coordinates $x$ and $y$ mimic the metric 
$g_{ab}$ and 
the scalar $\phi$, respectively, the functions $f(y)$ and  $w(y)$ represent  
$f(\phi)$ and $\omega(\phi)$, while $J$ represents the matter sources. Since 
we are interested in the purely gravitational sector, we will set $J$ to zero 
in most of the following. 

The coordinate $x$ is cyclic and the Euler-Lagrange equations 
$\frac{d}{dt}\left( \frac{\partial L}{\partial \dot{x}^i} \right) 
-\frac{\partial 
L}{\partial x^i}=0 $ ($i=1,2$) yield
\begin{eqnarray}
&& \dot{x}f(y)=C \;, \label{1001} \\
&&\nonumber \\
&& w(y)\ddot{y}+\frac{w'(y)  \dot{y}^2}{2} +\frac{f'(y)\dot{x}^2}{2}=0 
\;, 
\label{1002}
\end{eqnarray}
where a prime now denotes differentiation with respect to $y$ and $C$ is 
an 
arbitrary integration constant.

\noindent {\em i)} The analogue of a singularity of the first 
kind $f(\phi)=0$ in a domain is  $f(y)\equiv 0$ on an interval, which implies 
$C=0$ and
\be
w(y) \ddot{y}+ \frac{w'(y) \dot{y}^2}{2}=0 \;.
\ee
This equation admits the first integral
\be
\int_{y_0}^y dy' \, \sqrt{ |w(y')|} = C_1 \left( t-t_0\right) \;,
\ee
where $C_1$ and $t_0$ are integration constants. Let us consider now, for the 
sake of illustration, the choice $w(y)=y$ yielding the solution
\be
y(t)=C_2 \left( t-t_*\right)^{2/3} \;,
\ee
with $C_2$ and $t_*$ integration constants. Note that, because $f\equiv 0$, 
there is no equation for $x(t)$ and the dynamics of this variable are lost: 
the initial value problem is not well-posed because $x(t)$ can be assigned 
arbitrarily and is not determined uniquely by initial data $\left( x_0, 
\dot{x}_0 \right)$ at an initial time $t_0$.

\noindent {\em ii)}  Let us consider now the situation in which $f(y)$  
vanishes at isolated points $y_*$ mimicking the critical scalar field values 
$\phi_*$. Then, in the system~(\ref{1001}) and (\ref{1002}), either $C=0$ or 
else $\dot{x}\rightarrow \infty$ as $y\rightarrow y_*$. If $C\neq 0$, then 
$\dot{x}=C/f(y) \rightarrow \pm \infty$ as $f(y)\rightarrow  0^{\pm}$ and, 
therefore, also $x(t) \rightarrow \pm \infty$ or  the solution is not of 
class ${\cal C}^1$ and its derivative does not exist. In the first case, a 
barrier separates the regions $f(y)>0$ and $f(y)<0$, however special 
solutions which traverse the barrier $y=y_*$ can in principle exist. 

If $C=0$, an exceptional solution $x(t)=$const., $y(t)=C_2\left( 
t-t_*\right)^{2/3}$ passes through this barrier, however this corresponds to 
the special value $C=0$ and it disappears when $C\neq 0$.

\noindent {\em iii)} Let us consider now the case $w(y)\equiv 0 $ on an 
interval, corresponding to a singularity of the second kind $f_1(y)=0$ on a 
domain. Then, we are left with
\begin{eqnarray}
&& \dot{x}f(y)=C \;, \label{1003}\\
&&\nonumber \\
&& \dot{x}f'(y)=0 \;. \label{1004}
\end{eqnarray}
From eq.~(\ref{1004}), either $x(t)=$const. and then it must be $C=0$ with 
no equation left to determine $y(t)$, or   the equation $f'(y)=0$ is an 
algebraic (or trascendental, but not a 
differential) equation that determines constant values $y_*$ of $y$ (if it 
admits roots). Assuming that $f(y_*)\neq 0$, then $x_*(t)=\frac{C}{f(y_*)}\, 
t +x_0$. The solutions $\left( x_*(t), y_*(t) \right) $, if they exist, are 
the only ones and correspond to exceptional initial conditions and, in this 
sense, there are no dynamics for $y$.

\noindent {\em iv)} We can now consider the situation in which $w(y)$ 
vanishes 
at isolated points $y_c$, mimicking isolated singularities of the second kind 
$f_1(\phi_c)=0$. Consider, for example, the choice $w(y)=y$, $f(y)=y-1$, for 
which the system~(\ref{1001}) and (\ref{1002}) reduces to
\begin{eqnarray}
&& \dot{x}\left( y-1 \right)=C \;, \\
&& \nonumber \\
&& y\ddot{y}+\frac{\dot{y}^2}{2}+(y-1)\, \frac{\dot{x}^2}{2}=0 \;.
\end{eqnarray}
Assuming that $y$ is not identically unity, it is 
$ y\ddot{y}+\frac{\dot{y}^2}{2}+ \frac{ C^2}{2(y-1) }=0 $; at $y=0$ one has 
$\dot{y}_c=\pm C$ and one can not assign arbitrary initial conditions on the 
``hypersurface'' analogue $y=0$, but only the initial data 
$ \left( x_0,  \dot{x}_0, y_0,  \dot{y}_0 \right)= 
\left( x_0,  -C, 0,  \pm C \right) $ are allowed there, where $C$ and 
$x_0$ are arbitrary constants. The region  allowed to the 
dynamics in the four-dimensional space 
$ \left( x_0,  \dot{x}_0, y_0,  \dot{y}_0 \right) $ is only two-dimensional,  
due to the presence of the 
first integral~(\ref{1001}) and of the additional first integral 
\footnote{Eq.~(\ref{2ndfirstintegral})  can be easily derived from the field 
equations by multiplying eq.~(\ref{1002}) by $\dot{y}$  and integrating; 
vice-versa,  one verifies that  $d{\cal H}/dt=0$ by using the  equations 
of motion.}
\be  \label{2ndfirstintegral}
\frac{w(y)\dot{y}^2}{2}-\frac{C^2}{2f(y)}={\cal H}=\mbox{const.} 
\ee
If a solution attains the critical value $y=0$, it must 
assume the values 
$ \left( x_0,  \dot{x}_0, y_0,  \dot{y}_0 \right)= 
\left( x_0,  -C, 0,  \pm C \right) $ there, for which the ``energy'' 
${\cal H}$ can 
only take the values
\be
{\cal H}=\frac{C^2}{2} \left[ w(0)+1 \right] 
\ee
(where eq.~(\ref{1001}) has been used) everywhere along the orbits of the 
solutions.

\subsection{The analogue of the Einstein frame}

Let us consider again the toy model action~(\ref{toymodelaction}); the 
tranformation to the Einstein frame for scalar-tensor gravity is a change of 
variables modelled by the transformation $\left(x,y \right)\rightarrow \left( 
\xi, \eta \right)$ defined by
\begin{eqnarray}
d\xi & = & \sqrt{f(y)} \, dx \;, \label{analogconftran1}\\
&&\nonumber \\
d\eta & = & \sqrt{w(y)} \, dy \;. \label{analogconftran2}
\end{eqnarray}
In terms of these new variables, the action~(\ref{toymodelaction}) is 
rewritten in the canonical form 
\be
S=\int dt \left[ \frac{\dot{\xi}^2}{2} -\frac{\dot{\eta}^2}{2}-J\left( x(\xi, 
y) \right) \right] \;,
\ee
which mimicks the Einstein frame representation of the scalar-tensor action 
with the matter sources $J$ now depending on both the ``new metric'' $\xi$ 
and the ``scalar field'' ($y$, or $\eta$ through $y(\eta)$).   
Zeros of either $f(y)$ or $w(y)$ make the analog of the conformal 
transformation plus scalar field redefinition~(\ref{analogconftran1}), 
(\ref{analogconftran2}) ill-defined. Moreover, if $ f(y) \neq 0$ and only 
$w(y)$  vanishes, 
one can still consider an ``Einstein frame'' representation with the 
variables $\xi$ and $y$, in terms of which the action is simply 
\be 
S=\int dt \left[ \frac{\dot{\xi}^2}{2}-J(\xi, y) \right] \;.
\ee
It is clear that, similar to the case considered before for scalar-tensor 
gravity, there are no dynamics for the variable $y$, which can be assigned 
arbitrarily \footnote{This situation should not be confused with 
the $\omega=0$ Brans-Dicke equivalent of metric $f(R)$ gravity, for 
which there are indeed non-trivial dynamics even if no kinetic term for 
$\phi$ appears in the 
action. In fact, there, a dynamical equation containing $\Box \phi $ still 
exists to rule the evolution of $\phi$. It is instead Palatini $f(R)$ 
gravity which has no dynamical equation for $\phi$ because $\Box \phi $ 
completely disappears from the relevant equation  \cite{review}.}.  This is 
bad news if this variable plays a physical role because 
there are no equations to rule it and it can only  
be assigned from outside the theory, which is akin to invoking a 
miracle to produce any effect that one may desire and results 
in a complete loss of 
predictive power for the theory.

\subsection{Singular points of ODEs}

To conclude this section, we comment on the fact that, in the theory of 
ordinary differential equations (ODEs), it is rather common to encounter 
situations in which the phase space is divided into two disconnected regions, 
with only exceptional solutions, or  a restricted submanifold of 
solutions, 
crossing the  boundary between these two regions. Consider, for example, the 
ODE
\be
t^2 \ddot{y}-2y=0 \;,
\ee
which has $t=0$ as a regular singular point. Two linearly independent 
solutions are
\be
y_1(t)=t^2 \;, \;\;\;\;\;\;\;\;\;\; y_2(t)=\frac{1}{t} \;.
\ee
The first solution crosses undisturbed the  $t=0$ ``barrier'', while the 
second cannot (that is, $t=0$ is a barrier to at least some of the  
solutions). Consider also the third solution
in $\left( -\infty, 0 \right) \cup \left( 0, +\infty \right) $
\be
y_3(t)=\left\{
\begin{array}{l}
 0 \;\;\;\;\;\mbox{if}\;\;\;\; t\leq 0 \;,\\ 
 \\
 t^2 \;\;\;\;\;\mbox{if}\;\;\;\; t\geq 0 \;.
\end{array}
\right.
\ee
$y_3$ is continuous with its first derivative at $t=0$ (but the second 
derivative is not defined there). Now, $y_{1}(t)$ and $y_3(t)$ are linearly 
independent solutions which satisfy the same initial conditions $\left(y(0), 
\dot{y}(0) \right)=\left( 0,0 \right)$ at $t=0$. These two otherwise distinct 
solutions 
intersect at the origin of the phase space, which signals the breakdown of 
the 
initial value formulation at $t=0$. It is not surprising, therefore, that for 
the more complicated systems of partial differential equations ruling  
scalar-tensor theories, the Cauchy problem breaks down at the analogue of 
singular points of the equations. Depending on the particular form of the  
coupling functions $f(\phi)$ and $\omega(\phi)$, special 
solutions crossing the barrier may or may not exist. The  
situation 
in ODE theory in which no such special solution exists is exemplified by 
the equation
\be
t^2\ddot{y}+ 5t \dot{y} +3y=0 \;.
\ee
For $t>0$ and for $t<0$, two linearly independent solutions are $y_1(t)=1/t $ 
and $ y_2(t)=1/t^3$. No choice of the arbitrary constants $C_{1,2}$ in the 
general solution
\be
y(t)=\frac{C_1}{t}+ \frac{C_2}{t^3}
\ee
in $\left( -\infty, 0 \right) \cup \left( 0, +\infty \right) $ produces a 
solution crossing the barrier $t=0$.

\section{\label{sec:VII}VII.~Conclusions}

In principle, two kinds of ``singularities'' for the Cauchy 
problem are possible in   general scalar-tensor theories: those 
(``first kind'') at  which $f(\phi)=0$, and those (``second 
kind'') at which $f_1(\phi)=0$. Although statements that 
these should be rejected outright have been voiced in the 
literature \cite{FaresePolarski}, solutions 
corresponding to  critical values of the BD-like  scalar field of   both 
first  \cite{Linde, BarceloVisser, 
GunzigSaa, who?,  
FutamaseMaeda} and  
second kind \cite{Starobinsky, FutamaseMaeda, 
FutamaseRothmanMatzner, AbramoBGS, GunzigSaa} have been studied 
in the literature. Critical points of the second kind may 
appear benign when studied in a spatially homogeneous and 
isotropic FLRW universe, but 
they reveal their true nature of  geometrical singularities 
when analyzed in anisotropic  Bianchi models \cite{Starobinsky, 
FutamaseRothmanMatzner, AbramoBGS,GunzigSaa}. Here, 
following recent developments in the theory of the Cauchy 
problem of scalar-tensor gravity, we have shown that the latter is 
not well-posed at any of those critical points. The solutions are 
not unique and the physics becomes 
unpredictable. Physically, this is associated to a change in 
sign of the effective gravitational coupling~(\ref{20}), which 
diverges at both kinds of critical points. It seems that nature 
abhors such changes from attractive to repulsive gravity (and 
vice-versa) which, formally, only take place through a singularity of  
$G_{eff}$. This, however, says 
nothing 
about exotic forms of matter which can source repulsive gravity 
through the field equations, a completely  different and 
seemingly perfectly legitimate mechanism from the mathematical 
point of view (although the violation of all the energy 
conditions would certainly be questionable on  physical 
grounds).

To conclude, we remark that a possible cure for the problem of Palatini 
$f(R)$ 
gravity (already outlined in Refs.~\cite{BarausseSotiriouMiller, GunzigSaa, 
FutamaseRothmanMatzner}) could be the insertion into 
the gravitational action of terms that introduce higher order derivatives 
into the field equations. Then, the dropping out of $\Box \phi$ from the 
field equations  will 
be immaterial. However, unless such higher derivative terms appear in the 
Gauss-Bonnet combination, they will introduce ghost fields. A study of the 
initial value problem for these Gauss-Bonnet-corrected theories will be 
presented elsewhere.

\begin{acknowledgments}

VF acknowledges Marcelo Salgado for a useful comment. This work is supported 
by the Natural  Sciences 
and Engineering  Research Council of Canada.
\end{acknowledgments}

% Create the reference section using BibTeX:
%\bibliography{simplified}

\begin{thebibliography}{99}


\bibitem{SN} A.G. Riess {\em et al.}, {\em Astron. J.} 
{\bf 116} (1998) 1009 ; {\em Astron. J.} {\bf 118} (1999) 2668;
{\em Astrophys. J.} {\bf 560}  (2001) 49;
{\em Astrophys. J.} {\bf 607} (2004) 665;
S. Perlmutter {\em et al.}, {\em Nature} {\bf 391} (1998) 
51; 
{\em Astrophys. J.} {\bf 517} (1999) 565; 
J.L. Tonry {\em et al.}, {\em Astrophys. J.} {\bf 594} (2003) 
1; R.  Knop {\em et al.}, {\em Astrophys. J.} {\bf 598} 
(2003) 102;
B. Barris {\em et al.}, {\em Astrophys. J.} {\bf 602} 
(2004) 571;
A.G. Riess {\em et al.}, astro-ph/0611572.

\bibitem{extendedquintessence} 
T. Chiba, {\em Phys. Rev. D} {\bf 60} (1999) 083508; 
J.-P. Uzan,  {\em Phys. Rev. D}  {\bf 59} (1999) 123510;
F. Perrotta, C.  Baccigalupi,  and S. Matarrese,  {\em 
Phys. Rev. D} {\bf 61} (1999) 023507;
L. Amendola, {\em Mon. Not. R. Astr. Soc.} {\bf 312} (2000) 
521; R. de Ritis, A.A. Marino, C.  Rubano, and P. Scudellaro, 
 {\em Phys. Rev. D} {\bf 62} (2000) 043506;
X. Chen, R.J. Scherrer, and  G. Steigman, {\em Phys. Rev. 
D} {\bf 63} (2001) 123504; G. Esposito-Far\'{e}se, 
gr-qc/0011115;
A.A. Sen and R.T. Seshadri,  {\em Int. J. Mod. Phys. D} 
{\bf 12} ( 2000) 445;
C. Baccigalupi, S. Matarrese,  and F. Perrotta,  {\em 
Phys. Rev. D} {\bf 62} (2000) 123510;
Y. Fujii,  {\em Gravit. Cosmol.} {\bf 6} (2000) 107;
{\em Phys. Rev. D} {\bf 62} (2000) 044011;
O. Bertolami and  P.J. Martins,   {\em Phys. Rev. D} {\bf 
61} (2000) 064007; V. Faraoni, {\em Phys. Rev. D} {\bf 62} 
(2000) 023504; 
L. Amendola,  {\em Phys. Rev. D} {\bf 62} (2000) 043511;
{\em Phys. Rev. Lett.} {\bf 86} (2001) 196;
B. Boisseau, G. Esposito-Far\'{e}se, D. Polarski, and  
A.A. Starobinsky, {\em Phys. Rev. Lett.}  {\bf 85} (2000) 2236;
M.C. Bento, O. Bertolami, and N.C. Santos,   {\em Phys. 
Rev. D} {\bf 65} (2002) 067301;
T. Chiba, {\em Phys. Rev. D} {\bf 64} (2001) 103503;
A. Riazuelo and J.-P. Uzan,  {\em Phys. Rev. D} {\bf 62} (2000) 
083506;
N. Banerjee and D. Pavon,  {\em Class. Quantum Grav.} {\bf 
18} (2001) 593;
V. Faraoni, {\em Int. J. Mod. Phys. D} {\bf 11} (2002) 471;
D.F. Torres,  {\em Phys. Rev. D} {\bf 66} (2002) 043522;
F. Perrotta and C. Baccigalupi, astro-ph/0205217;
S. Nojiri and S.D. Odintsov,  {\em Phys. Rev. D} {\bf 
68} (2003)  123512; 
E. Elizalde {\em et al.},  {\em Phys. Lett. B} {\bf 574}  
(2003) 1; 
S. Nojiri  and  S.D. Odintsov,  {\em Phys. Rev. D} {\bf 
70} (2004) 103522; {\em Phys. Lett. B} {\bf 595} (2004) 
1; {\em Gen. Rel. Grav.} {\bf 38} (2006) 1285; E. Elizalde, S. 
Nojiri, and S.D. 
Odintsov, {\em Phys. Rev. D} {\bf 
70} (2004) 043539;    S. Matarrese, C. Baccigalupi,  and  F. Perrotta, 
astro-ph/0403480; V. Faraoni, {\em Phys. Rev. D} {\bf 68} 
(2003) 063508;  {\bf 69} (2004) 123520;  {\bf 70} 
(2004) 044037; {\em Ann. Phys. (NY)} {\bf 317} 
(2005) 366;  S. Capozziello, S. Nojiri, and S.D. Odintsov, {\em 
Phys. Lett. B} {\bf 634} (2006) 93; 
V. Faraoni, M.N. Jensen, and S. Theuerkauf, 
{\em Class. Quantum Grav.} {\bf 23} (2006) 4215; V. Faraoni and  
M.N. Jensen, {\em Class. Quantum Grav.} {\bf 23} (2006) 3005; 
R. Catena, M. Pietroni, and L. Scarabello, {\em J. Phys. A} 
{\bf 40} (2007) 6883; {\em Phjys. Rev. D} {\bf 76} (2007) 
084039;  E. Elizalde, S. Nojiri, S.D. Odintsov, D. Saez, and V. 
Faraoni, {\em Phys. Rev. D} {\bf 77} (2008) 106005.

\bibitem{CCT} S. Capozziello, S. Carloni, and A. 
Troisi,  astro-ph/0303041. 

\bibitem{CDDT} S.M. Carroll, V. Duvvuri, M. Trodden, and M.S.  
Turner, {\em Phys. Rev. D} {\bf 70} (2004) 043528.

\bibitem{review} T.P. Sotiriou and V. Faraoni, arXiv:0805.1726. 

\bibitem{Wald} R.M. Wald, {\em General Relativity} (Chicago 
University Press, Chicago, 1984).

\bibitem{Vollick} 
D.N. Vollick, {\em Phys. Rev. D} {\bf 68} (2003) 063510.

\bibitem{SotiriouLiberati} T.P. Sotiriou  and S. Liberati, {\em  
Ann.  Phys. (NY)} {\bf 322} (2007)  935; T.P. Sotiriou, {\em 
Class.  Quantum  Grav.} {\bf 23} (2006) 5117;  arXiv:0710.4438.


\bibitem{STequivalence} P.W. Higgs, {\em Nuovo Cimento} 
{\bf 11} (1959) 816;   P. Teyssandier and P. Tourrenc, 
{\em J. Math. Phys.} {\bf 24} (1983) 2793; B. Whitt, {\em 
Phys. Lett. B} {\bf 145} (1984) 176;  J.D. Barrow and S. 
Cotsakis, {\em Phys. Lett. B} {\bf 214} (1988) 515; J.D. 
Barrow,  {\em Nucl. Phys. B} {\bf 296} (1988) 697;  D. 
Wands,  {\em Class. Quantum Grav.} {\bf 11} (1994) 269; T. 
Chiba, {\em Phys. Lett. B} {\bf 575} (2003) 1.

\bibitem{BD} C.H. Brans  and R.H. Dicke, {\em 
Phys. Rev.}  {\bf 124} (1961) 925. 

\bibitem{Dicke} R.H. Dicke, {\em Phys. Rev.} {\bf 125} (1962) 
2163.

\bibitem{ST}  P.G. Bergmann, {\em Int. J. Theor. Phys.}  
{\bf 1}  (1968) 25;
R.V. Wagoner, {\em Phys. Rev. D} {\bf 1} (1970) 3209;
K. Nordvedt, {\em Astrophys. J.} {\bf 161} (1970) 1059. 

\bibitem{bosonicstring} C.G. Callan, D. Friedan, E.J. Martinez, 
and M.J. Perry, {\em Nucl. Phys. B} {\bf 262} (1985) 593; 
E.S. Fradkin and A.A. Tseytlin, {\em Nucl. Phys. B} {\bf 261} 
 (1985) 1.

\bibitem {Duffetal} M.J. Duff, R.R. Khuri, and J.X. Lu, {\em 
Phys. Rep.} {\bf 259} (1995) 213.

\bibitem{mybook} V. Faraoni, {\em Cosmology in Scalar-Tensor 
Gravity} (Kluwer Academic, Dordrecht, 2004).

\bibitem{FujiiMaeda} Y. Fujii and K. Maeda, {\em The 
Scalar-Tensor Theory of Gravity} (CUP, Cambridge, England, 
2003).

\bibitem{Salgado} M. Salgado, {\em Class. Quantum Grav.} {\bf 
23} (2008) 4719. 

\bibitem{HawkingEllis} S.W. Hawking and G.F.R. Ellis, {\em The 
Large Scale Structure of Spacetime} (CUP,  Cambridge, England, 
1973).

\bibitem{LanahanFaraoni} N. Lanahan-Tremblay and V. Faraoni, 
{\em Class. Quantum Grav.} {\bf 24} (2007) 5667.

\bibitem{Salgadoetal} M. Salgado, D. Martinez-del Rio, M. 
Alcubierre, and D. Nunez,  arXiv:08001.2372.

\bibitem{Flanagan} E.E. Flanagan, {\em Class. Quantum Grav.} 
{\bf 21} (2004) 3817.

\bibitem{FaraoniNadeau} V. Faraoni and S. Nadeau, {\em Phys. 
Rev. D} {\bf 75} (2007) 023501.

\bibitem{FaresePolarski} G. Esposito-Far\'ese and D. Polarski, 
{\em Phys. Rev. D} {\bf 63} (2001) 063504.

\bibitem{FutamaseMaeda} T. Futamase and K.I. Maeda, 
{\em Phys. Rev. D} {\bf 39} (1989) 399.

\bibitem{FutamaseRothmanMatzner} 
T. Futamase, T. Rothman, and R. Matzner, {\em Phys. 
Rev. D} {\bf 39} (1989) 405.

\bibitem{Hosotani} Y. Hosotani, {\em Phys. Rev. D} {\bf 32} 
(1985) 1949.

\bibitem{Linde} A. Linde, {\em JETP Lett.} {\bf 30} (1980) 447.

\bibitem{Starobinsky} A.A. Starobinsky, {\em Sov. Astron. 
Lett.} {\bf 7} (1981) 36.

\bibitem{BronnikovKireev} K.A. Bronnikov and Yu.N. Kireev, 
{\em Phys. Lett. A} {\bf 67} (1978) 95.

\bibitem{BarceloVisser} C. Barcelo and M. Visser, {\em Class. 
Quantum Grav.} {\bf 17} (2000) 3843.

\bibitem{AbramoBGS} L.R. Abramo, L. Brenig, E. Gunzig, and A. 
Saa, {\em Phys. Rev. D} {\bf 67} (2003) 027301; {\em 
Int. J. Theor. Phys.} {\bf 42} (2003) 1145.

\bibitem{VFsingularities} V. Faraoni, {\em Phys. Rev. D} {\bf 
70} (2004) 047301.

\bibitem{LiddleWands} A.R. Liddle and D. Wands, {\em Phys. Rev. 
D} {\bf 45} (1992) 2665.

\bibitem{TorresVucetich} D.F. Torres and H. Vucetich, {\em 
Phys. Rev. D} {\bf 54} (1996) 7373.

\bibitem{Nordvedt} K. Nodvedt, {\em Phys. Rev. D} {\bf 169} 
(1968) 1017.

\bibitem{Boisseauetal} B. Boisseau, G. Esposito-Farese, D. 
Polarski, and A.A. Starobinsky, {\em Phys. Rev. Lett.} {\bf 
85} (2000) 2236.

\bibitem{who?} E. Gunzig, A. Saa, L. Brenig, V. Faraoni, T.M. 
Rocha Filho, and A. Figueiredo, {\em Phys. Rev. D} {\bf 63} 
(2001) 067301; A. Saa, E. Gunzig, L. Brenig, V. Faraoni, T.M. 
Rocha Filho, and A. Figueiredo, {\em Int. J. Theor. Phys.} {\bf 
40} (2001) 2295; E. Gunzig, V. Faraoni, A. Figueiredo, T.M. 
Rocha Filho and L. Brenig, {\em Class. Quantum Grav.} {\bf 17} 
(2000) 1783.

\bibitem{superquintessence} V. Faraoni, {\em Int. J. Mod. 
Phys. D} {\bf 11} (2002) 471.

\bibitem{DamourFarese} T. Damour and G. Esposito-Far\`ese, {\em 
Class. Quantum Grav.} {\bf 9} (1992) 2093.

\bibitem{Gurevich} L.E. Gurevich, A.M. Finkelstein, and V.A. 
Ruban, {\em Astrophys. Space Sci.} {\bf 22} (1973) 231.

\bibitem{Pollock} M.D. Pollock, {\em Phys. Lett. B} {\bf 103} 
(1982) 386.

\bibitem{GunzigNardone} E. Gunzig and P. Nardone, 
{\em Phys. Lett. B} {\bf 134} (1984) 412.

\bibitem{Novello} M. Novello, {\em Phys. Lett. A} {\bf 90} 
(1982) 347.
 
\bibitem{Bronnikov} K.A. Bronnikov and M.S. Chernakova, {\em 
Gravit. Cosmol.} {\bf 11} (2005) 305.

\bibitem{BarausseSotiriouMiller}  E. Barausse, T.P. Sotiriou, 
and J. Miller, {\em Class. Quantum  Grav.} {\bf 25} (2008) 
062001;  arXiv:0712.1141;  arXiv:0801.4852; see also B. Li, 
D.F. Mota, and D.J. Shaw, arXiv:0805.3428.

\bibitem{CockeCohen} W.J. Cocke and J.M. Cohen, {\em J. Math. 
Phys.} {\bf 9} (1968) 971.

\bibitem{Noakes} D.R. Noakes, {\em J. Math. Phys.} {\bf 24} 
(1983) 1846.

\bibitem{York} J. York, in {\em Sources of Gravitational 
Radiation}, L. Smarr ed. (CUP, Cambridge, England, 1979).

\bibitem{BellucciFaraoni} S. Bellucci and V. Faraoni, {\em Nucl. 
Phys. B} {\bf 640} (2002) 453.

\bibitem{GunzigSaa} E. Gunzig and A. Saa, {\em Int. J. Theor. 
Phys.} {\bf 43} (2004) 575.


\end{thebibliography}

\end{document}